\documentclass[twocolumn,prl,aps,superbib,tightenlines,floatfix]{revtex4}
\usepackage{amsfonts,amsmath,amssymb,amsthm,bm}
\usepackage{graphicx}
\begin{document}
\bibliographystyle{apsrev}
\title{Alkanethiol-Based Single-Molecule Transistors}
\author{Chun-Lan Ma}
\author{Diu Nghiem}
\author{Yu-Chang Chen}
\email{yuchangchen@mail.nctu.edu.tw}
\affiliation{Department of Electrophysics, National Chiao Tung University, 1001 Ta Hsueh Road,
Hsinchu 30010, Taiwan  }
\begin{abstract}

We have investigated the transport properties of alkanethiol molecules in the two-terminal and three-terminal junctions by using first-principles approaches. We observe that states around the Fermi levels are introduced in the amino-substituted butanethiol junction. It leads to a sharp increase of the current which is credited to the resonant tunneling. The current-voltage characteristics suggest that the amino-substituted butanethiol molecular junction may be a promising candidate for field-effect transistors.

\end{abstract}
\maketitle

The purpose of molecular electronics is to develop electronic devices based on molecules. The concern with metal-molecule-metal (m-M-m) tunnel junctions has been growing due to their potential applications in subminiature devices. Substantial progress towards the goal has been achieved in various molecule systems. Over the past decade, a considerable number of studies have been conducted on the transport properties of alkane junctions.
\cite{Tao,Wold1,Cui,Wold2,Mizutani,Wold3,Wang,Beebe,Bumm,Sachs,Engelkes,Kaun,Smalley,Slowinski1} One of the reasons for this is that alkanethiols form a robust self-assembled monolayer (SAM) on gold surface.
\cite{Tao,Wold1,Cui,Wold2,Mizutani,Wold3,Wang,Beebe,Bumm,Sachs,Engelkes} Nevertheless, device properties have not yet been observed in the current-voltage characteristics of alkanethiol junctions. \cite{Tao,Wold1,Cui,Wold2,Mizutani,Wold3,Wang,Beebe,Sachs,Engelkes,Kaun,Mowbray} We explore a possible way to produce the current-voltage characteristics in alkane junctions that are applicable to device design. In this regard, this letter reports a theoretical study on the conductance in an alkanthiol-based single-molecule junction as a function of source-drain bias ($V_{\texttt{SD}}$) and gate voltage ($V_{\texttt{G}}$). The results clearly show that conductance can be modulated significantly by the source-drain biases and gate voltages in the amino-substituted butanethiol junction, which could be a promising candidate for field-effect transistors.
%%In addition, the butanethiols form a SAM on the Au surface that could be beneficial to technology attempting to integrate a great amount of transistors in a
%%small region.

\begin{figure}
\includegraphics[width=9cm]{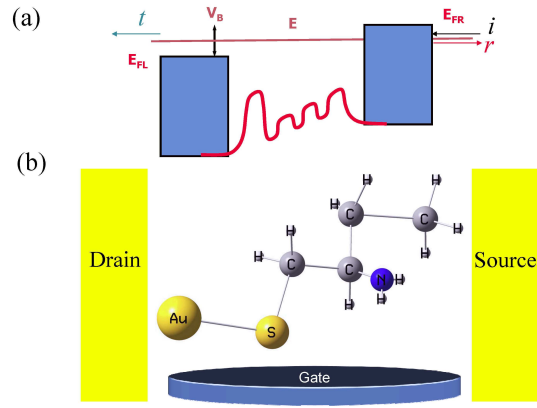}
\caption{(color online) (a) Energy diagram of two metal electrodes kept at a certain source-drain voltage. The left electrode is positively biased. (b) Schematic of the three-terminal junction used in the present study. The (amino-substituted) butanethiol molecule is sandwiched between the source and drain electrodes.}
\label{fig1}
\end{figure}

Alkanethiols [CH$_{3}$(CH$_{2}$)$_{n-1}$SH, denoted as C$_{n}$] are a good illustration of reproducible junctions that can be fabricated. \cite{Tao} It has been established that non-resonant tunneling is the main conduction mechanism in alkane junctions as the Fermi levels of the two electrodes lie within the large HOMO-LUMO gap (HOMO: highest occupied molecular orbital; LUMO: lowest unoccupied molecular orbital). \cite{Wold1,Cui,Wold2,Bumm} Non-resonant tunneling leads
to small and linear current-voltage characteristics over a wide range of biases unfavorable for device application. In order to modulate
the conductance with effective control, we explore the amino-substituted butanethiol (C$_{4}$) junction from three perspectives. Firstly, it
is possible that additional molecular orbitals between the HOMO-LUMO gap are produced when -NH$_{2}$ is substituted for -H in the bridging butanethiol,
the result of which may lead to resonant tunneling in the amino-substituted junction. Secondly, it has been reported that enhanced gate-controlled
conductance and NDR have been observed in some junctions containing amino or nitro moieties, \cite{Chen} but those effects are unknown in the
amino-substituted alkanethiol junction.  Thirdly, -NH$_{2}$ may be induced into the junction in the process of sample preparation, which means that -H can be substituted
by -NH$_{2}$ in the butanethiol junction. \cite{Venkataraman} It is for these reasons that we intend to explore the possibility of current-voltage characteristics in this system applicable to the design 
of device .

\begin{figure}
\includegraphics[width=9cm]{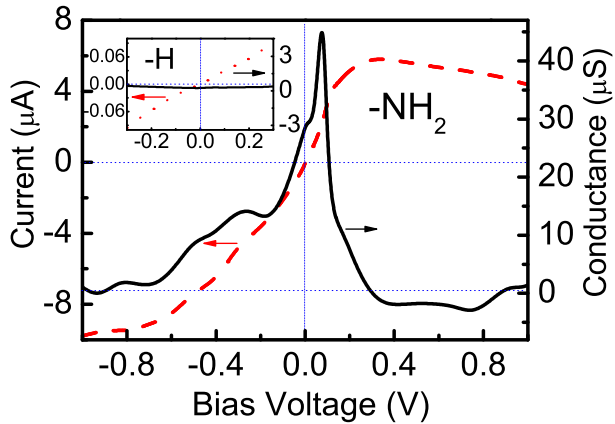}
\caption{(color online)  The current (left axis) and differential conductance (right axis) as a function of $V_{\texttt{SD}}$ in the
amino-substituted butanethiol junction (main graph) and the unsubstituted butanethiol junction (the inset).}
\label{fig2}
\end{figure}

In this letter, we investigate the dependence of conductance on the source-drain bias and gate voltage before and after amino substitution in
the butanethiol molecular junction. The physical system of interest to us consists of a (amino-substituted) butanethiol between two bulk electrodes which is kept at a certain source-drain bias. In the two-terminal system, the Fermi level in the right/left electrodes is determined by filling the conduction band with the valence electrons in the bulk metal electrode described by jellium model ($r_{s}\approx3$). When the source-drain bias is applied, the chemical potentials of the left and right electrodes are shifted, i.e., the source drain bias is $V_{SD}=(E_{FR}-E_{FL})/e$ as shown in Figure$~\ref{fig1}$(a). The gate field is introduced as a capacitor composed of two parallel circular charged disks separated at a certain distance from each other. The axis of the capacitor is perpendicular to the transport direction. One plate is placed close to the molecule while the other plate, placed far away from the molecule, is set to be the zero reference field (Figure$~\ref{fig1}$(b)).  The $I$-$V_{\texttt{SD}}$ in a two-terminal and $I$-$V_{\texttt{G}}$ characteristics in a three-terminal geometry have been studied by using the first principles approaches, where the stationary scattering wave functions of the whole system are calculated by solving the Lippmann-Schwinger equation iteratively until self-consistency is obtained. \cite{Di Ventra1,Lang,Di Ventra2,Kohn} The exchange and correlation energies are included in the density functional formalism within the local density approximation. The wave functions are then applied to calculate steady state current. The results of the calculations show that the amino-substituted junction may be a promising single-molecular transistor.

%%The alkanethiol molecule within the junction may undergo irreversible changes at large biases. \cite{Mizutani}
We calculate the $I$-$V_{\texttt{SD}}$ characteristics of unsubstituted butanethiol junction (the inset of Figure$~\ref{fig2}$) in a stable region. \cite{Mizutani,Wold3} Although asymmetric binding in the junction typically leads to asymmetric $I$-$V_{\texttt{SD}}$ curve, such asymmetric characteristic is negligibly small particularly in the alkanethiol junction. The alkanethiol junction has a very large HOMO-LUMO gap and the Fermi levels lie in the middle of the gap. It is for this reason that the electrical response under positive and negative biases is quite linear. Similar to previous reports, \cite{Cui,Wold3,Wang,Beebe,Kaun} the current-voltage characteristics of the unsubstituted butanethiol junction is marked by a linear and symmetric relation. The findings are in agreement with the $I$-$V_{\texttt{SD}}$ characteristics explained by non-resonant tunneling mechanism. \cite{Wold1,Wang,Bumm,Sachs,Smalley,Slowinski1,Magoga,Slowinski} The low-bias resistance is around $3.3\times10^{6}~\Omega$, which agrees well
with the reported experimental \cite{Wold1,Wold3,Beebe,Engelkes} and theoretical results. \cite{Kaun}

\begin{figure}
\includegraphics[width=9cm]{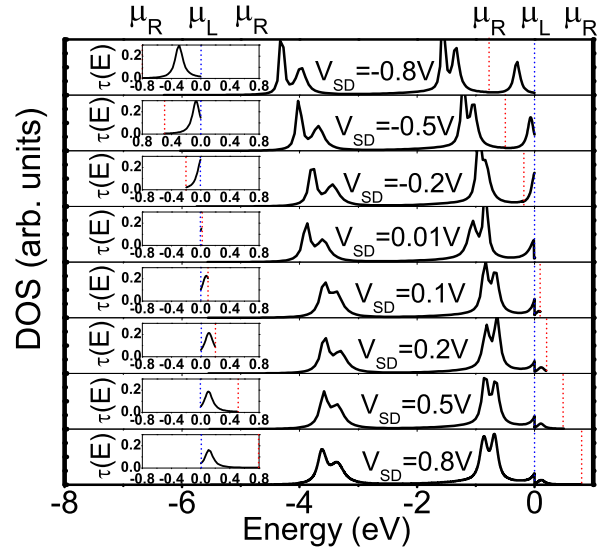}
\caption{(color online) The density of states and the transmission
probabilities $\tau$ (inset) for various source-drain
biases ($V_{\texttt{SD}}=-0.8$, $-0.5$, $-0.2$, $0.01$, $0.1$, $0.2$, $0.5$ and $0.8~$V) in
the amino-substituted butanethiol junction. The left Fermi
level $E_{\texttt{FL}}$ (blue dotted lines) is set to be
the zero of energy. The right Fermi level $E_{\texttt{FR}}$ (red dotted line)
defines $V_{\texttt{SD}}=(E_{\texttt{FR}}-E_{\texttt{FL}})/e$.}
\label{fig3}
\end{figure}

In order to investigate the effect of the substitution, we present current (\emph{I}) and differential conductance ($dI/dV_{\texttt{SD}}$) as a
function of source-drain bias ($V_{\texttt{SD}}$) for amino-substituted butanethiol junction in Figure$~\ref{fig2}$.
The figure indicates that the effects of the amino substitution in butanethiol junctions are pronounced. At around zero bias, the magnitude of
the current changes significantly in contrast to the unsubstituted case. AS the magnitude of $V_{\texttt{SD}}$ increases, there is a sharp increase in the magnitude of the current until it reaches saturation at around $V_{\texttt{SD}}=-0.8~$V and $V_{\texttt{SD}}=0.3~$V. A peak (up to $46.3~\mu$S) in the differential conductance is found at a bias around $0.1~$V. For positive bias, the current slightly decreases as the bias increases for $V_{\texttt{SD}}\geq0.3~$V.

\begin{figure}
\includegraphics[width=9cm]{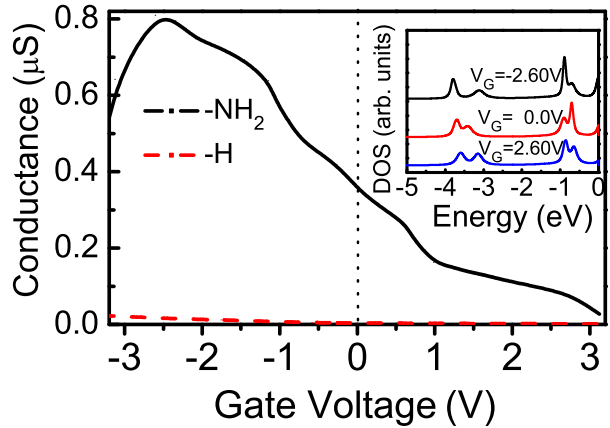}
\caption{(color online) The source-drain conductance as a function of gate voltage in a three-terminal geometry for an unsubstituted (red
dashed line) and an amino-substituted (black solid line) butanethiol junction, where the source-drain bias is fixed at $0.01~$V. The inset shows the density of states for an amino-substituted butanethiol junction for $V_{\texttt{G}} =-2.60$, $0.0$, and $2.60~$V, respectively ($V_{SD}=0.01~$V).}
\label{fig4}
\end{figure}

Let us now attempt to explain the above observations via a series of density of states (DOSs) and the transmission probabilities for various source-drain biases. As shown in Figure$~\ref{fig3}$, the responses of the density of states to the negative and positive biases seem to be very different in the amino-substituted butanethiol junctions. It leads to an asymmetric $I-V_{SD}$ characteristic of the device. In the regime of negative biases, we observe that the shapes of transmission probabilities and DOSs are in accordance. The LUMO peak created by the amino substitution lies slightly above the Fermi levels in the linear response regime. It is for this reason that the conductance can be significantly modulated by small source-drain biases. As the current-carrying window formed between E$_{\texttt{FL}}$ and E$_{\texttt{FR}}$ becomes wider with increasing magnitude of negative bias, more states from LUMO are gradually included. It provides evidence for the existence of resonant tunneling in the amino substituted butanethiol junctions in sharp contrast to the non-resonant tunneling in the unsubstituted system. The current seems to reach saturation at $V_{SD}=-0.8$~V, where the LUMO states are entirely included in the current-carrying energy windows. In the regime of positive biases, the DOSs and the transmission probabilities seem inert to change the shapes and the positions relative to the left Fermi level by the source-drain biases. Thus, the $I-V_{SD}$ curve shown in Figure$~\ref{fig2}$ is asymmetric. The response of the DOSs upon the positive biases is inert. This property may be related to the states in the current-carrying window being developed into a resonant peak similar to what is found in the elongated silicon point contactIt. \cite{Di Ventra3} For $V_{SD}>0.2$~V, the current tends to reach saturation since the current-carrying windows covers the entire resonant peak. In this regime, weak NDR effects are observed possibly due to a leak of states from the current-carrying windows or possible numeric instability.

As three-terminal field-effect transistor-like devices are highly desirable, we also investigate source-drain conductance as a function of gate
voltage when a small source-drain bias is applied. In Figure$~\ref{fig4}$, we compare the conductance of the butanethiol junction before and
after amino substitution. In the unsubstituted system, the small and featureless conductance stems from non-resonant tunneling, realizing that
the Fermi levels lie between the HOMO and LUMO gap. After amino substitution, as the gate voltage varies from $-2.60~$V to $3.12~$V, the
conductance decreases up to 30 times from $0.81~$G$_{0}$ to $0.027~$G$_{0}$. To explain why gate voltage can significantly modulate the
source-drain conductance, we examine the DOSs for various gate voltages as shown in the inset of Figure$~\ref{fig4}$. At zero gate voltage, the energies between two Fermi levels open a current-carrying window at the left wing of the LUMO spectrum, which induces resonant tunneling and corresponds to a large conductance compared with the unsubstituted system. The positive (negative) gate voltage shifts the broadened LUMO peak towards
higher (lower) energies. Negative gate voltages strengthen the resonant tunneling by shifting the position of the LUMO peak towards the
current-carrying window. At a gate voltage of $-2.60~$V, the central peak of LUMO is between the left and right Fermi levels, and the
conductance reaches a maximum. Conversely, a positive gate voltage shifts the position of the LUMO peak away from the current-carrying window,
and therefore, the source-drain conductance decreases.

Considering the density of states for an unsubstituted butanethiol junction, we observe that Fermi levels lie between the HOMO-LUMO gap. Thus,
the effect of gate voltage is irrelevant to the conductance of the system. Nevertheless, when -H in the molecule is substituted with -NH$_{2}$,
the gate voltage can easily modulate the current and therefore the conductance. All these characteristics suggest that the
amino-substituted butanethiol junction could be a field-effect transistor. Transistor-like behavior has been found in carbon
nanotubes \cite{Wildoer,Tans1,Tans2,Martel} and phenyldithiolate molecule junction, \cite{Yang} but it has never been found in alkanethiol
junctions. As the parent material (alkanethiols) can form a self-assembled monolayer, the amino-substituted junction might have the potential to
enable large-scale integration of single molecule transistors into a small region for miniaturization devices.

The authors are grateful to Professors Massimiliano Di Ventra, Prof. Zhongqin Yang for their helpful discussions. This work is supported by Taiwan NSC97-2112-M-009-011-MY3, MOE ATU, NCTS, and computing resources from NCHC.

\end{document}